\documentstyle[aaspp4]{article}

\def\msolar{ {M_{\odot}} }

\begin{document}

\title{Warm Neutral Gas at Redshift 3.4}

\author{F. H. Briggs\altaffilmark{1},  E. Brinks\altaffilmark{2,3}, 
and A. M. Wolfe\altaffilmark{4}}
\altaffiltext{1}{Kapteyn Astronomical Institute, Postbus 800,
9700 AV Groningen, The Netherlands}
\altaffiltext{2}{National Radio Astronomy Observatory, PO Box O,
Socorro, NM 87801-0387}
\altaffiltext{3}{currently on leave at Departamento de Astronom\'\i a,
 Universidad de Guanajuato, M\'exico}
\altaffiltext{4}{Center for Astrophysics and Space Science, C-011,
University of California at San Diego, La Jolla, CA 92093-0111}


\begin{abstract}

Radio spectroscopy at 323 MHz using  the 
Arecibo Telescope\footnote{
The National Astronomy and Ionosphere Center is operated
by Cornell University under contract with the National Science
Foundation}
and the VLA\footnote{The National Radio
Astronomy Observatory is a facility of the National Science Foundation
operated under cooperative agreement by Associated Universities,
Inc.}
has produced a tentative detection of the  
21--cm line of neutral hydrogen at $z_{abs}=3.38716\pm0.00007$ 
in absorption against the radio continuum of the QSO MG0201+113
($z_{em} = 3.61$). This redshift roughly 
agrees with one determined by the Westerbork
Synthesis Radio Telescope and reported earlier; however the observations
at different telescopes 
produce different results for width and optical depth of the line.

If the detection holds, it provides evidence for a high column density of
neutral gas that is confined to a dynamically cold layer with 
velocity dispersion $\sim$10 km~s$^{-1}$.  
Although the interpretation is uncertain due to a lack of detailed
knowledge of the
extended radio structure of the background quasar and the relative quantities
of neutral gas in the cold and turbulent components,
the observations specify  high spin temperatures, $T_s \geq
1000$ K for both this 21--cm line absorbing cloud and the turbulent component, which together have
$N_{HI} \approx 10^{21.4}$cm$^{-2}$.
Results of optical spectroscopy
require the additional presence of metal-enriched
clouds of still broader velocity dispersion than the 21--cm line, 
creating a picture which
is consistent with this system  being a young
disk galaxy that is sufficiently evolved by $z=3.4$ to have collapsed to a
flattened system and produced a population of stars that have aged 
to  pollute a turbulent halo.

The observations  constrain the neutral gas mass of a possible 21--cm line
emitter associated with the intervening absorber
to be less than ${\sim}(\Delta V/200$ km~s$^{-1}) 10^{13}\msolar$ for
velocity widths $\Delta V$ (FWHM) in the range 200 to 1200 km~s$^{-1}$ 
($H_o=100$ km~s$^{-1}$Mpc$^{-1}$,
$\Omega_o=1$).

\end{abstract}


\section{Introduction}

The application of
research into QSO absorption-lines to the field of galaxy formation
relies on the interpretation
that ordinary galaxies  are likely to intervene at random 
  along the line of sight to cosmologically
distant background QSOs. This central hypothesis is verified
by statistical tests for the ``metal-line'' and  ``damped Lyman~$\alpha$''
systems, which are identified with galaxian halos and HI-disks, respectively
(cf. Sargent \markcite{S1} 1988, Wolfe \markcite{Wo3} 1988).  
At high redshift, where ordinary galaxies are too distant to be viewed in
their own emitted light, QSO absorption-lines provide the only clues
to the state of evolution for the bulk of the galaxy population.
Clearly, objects with large
gaseous cross sections and high space densities are the systems most 
frequently encountered. 
This paper shows that the MG0201+113 $z_{abs}\approx 3.4$ absorption
system, which is
a typical high redshift system selected on the basis of
strong, damped Lyman~$\alpha$ absorption (cf. Lanzetta \markcite{L2} 1993), 
has already collapsed to a
low velocity dispersion layer.  Analysis of the associated metal lines
reveals an additional gaseous component of broader velocity spread that
can be identified with a turbulent halo; in these respects, the findings
are consistent with the 21--cm absorption systems
discovered at lower redshifts (Briggs \& Wolfe 1983, Wolfe 1986, Briggs 1988).

The Arecibo and VLA observations reported here for MG0201+113
constrain the optical
depth of the 21--cm absorption-line to be significantly less than the value 
reported by Baum et al \markcite{B1}
(1994) and de Bruyn et al \markcite{B2} (1996) for this system, 
although the measured redshifts are in good agreement.  Deduction
of high spin temperature in the in the absorber are consisted with the
findings of de Bruyn et al (1996) and with determinations for other
redshifted 21--cm absorption lines (Wolfe 1986, Briggs 1988, 
Cohen et al \markcite{Co1} 1994,
Steidel et al \markcite{St1} 1994,
Carilli 1995, Carilli et al 1996).

The new observations also place constraints on emission in the 21--cm line
from a massive HI-rich structure, such as a ``pancake'' (Zel'dovich 
\markcite{Z1} 1970)
or collapsing galaxy proto-cluster, which might be associated with the
damped Lyman~$\alpha$ absorber (Wolfe \markcite{Wo4} 1993).
Surveys to detect neutral hydrogen in emission at high redshift
through the observation of the 21--cm line
are now probing substantial volumes of space; Wieringa, de Bruyn 
and Katgert \markcite{W1} (1992) present a
review of recent surveys.  Taramopoulos, Briggs and Turnshek \markcite{T1}
(1994)
conducted a recent VLA study of the environment of the $z\approx 3.4$
damped Lyman~$\alpha$ absorber in the spectrum of Q0000-263 to perform
a similar search for 21--cm emission.

\section{The Arecibo Observations}

Radio spectroscopy of a narrow band around the expected sky frequency
of the redshifted 21--cm line for MG0201+113
was performed with the 318 MHz system
of the Arecibo Telescope in sessions 
on October 7, 8, 18, 20, 26 and November 19, 23, 24, 25, 1993. 
The details of the instrumentation are given by Briggs, Sorar and
Taramopoulos \markcite{Br4}
(1993). The only significant change since then has been the
improvement of the RF preamplifier so that system temperatures as low
as 65~K were measured on regions of sky far from the Galactic plane.
A 2.5 MHz bandwidth signal from the single linear polarization available
with the 318 MHz feed was fed to two parallel, 1024 channel banks of the 
Arecibo Correlator; the pair of resulting spectra were averaged to
obtain a slight improvement in signal-to-noise ratio resulting from
the different clip levels between the two banks, which were operated
in the three-level mode.  After a single Hanning smoothing, the 
spectra have a resolution of 4.88 kHz.

The observations were made in the ``total-power mode,'' in which an
on-source integration of 5 one-minute records was followed by a similar set
of 5 records taken off-source at 
sky coordinates that are chosen 
to force the telescope to track through the same antenna coordinates as
it did during the on-source scan.  In order to calibrate the system gain
as a function of frequency, similar ON-OFF scans were
obtained for the bright radio source 0320+0523. The continuum spectrum
of 0320+0523 is nearly flat for frequencies below 400 MHz so the flux density
7.63$\pm$0.18~Jy listed in the Texas Survey at 365 MHz (c.f. Douglas et al.
\markcite{D3} 1980) was adopted to
fix the flux scale of the observation.
Subtraction of each one-minute off-source spectrum from its
corresponding on-source spectrum produced a set of difference spectra
for both the calibration source and the program source, MG0201+113, and
these difference spectra were carefully inspected to identify
anomalous behavior. After excluding
difference spectra that were corrupted by
strong interference, the difference
spectra for the calibration source were averaged to form 
gain functions for division into the difference spectra for the program sources
to correct for the spectral dependence of the 
receiving system gain. A total of 270 one minute difference spectra
for MG0201+113
remained after editing; 130 were discarded.

{
\plotone{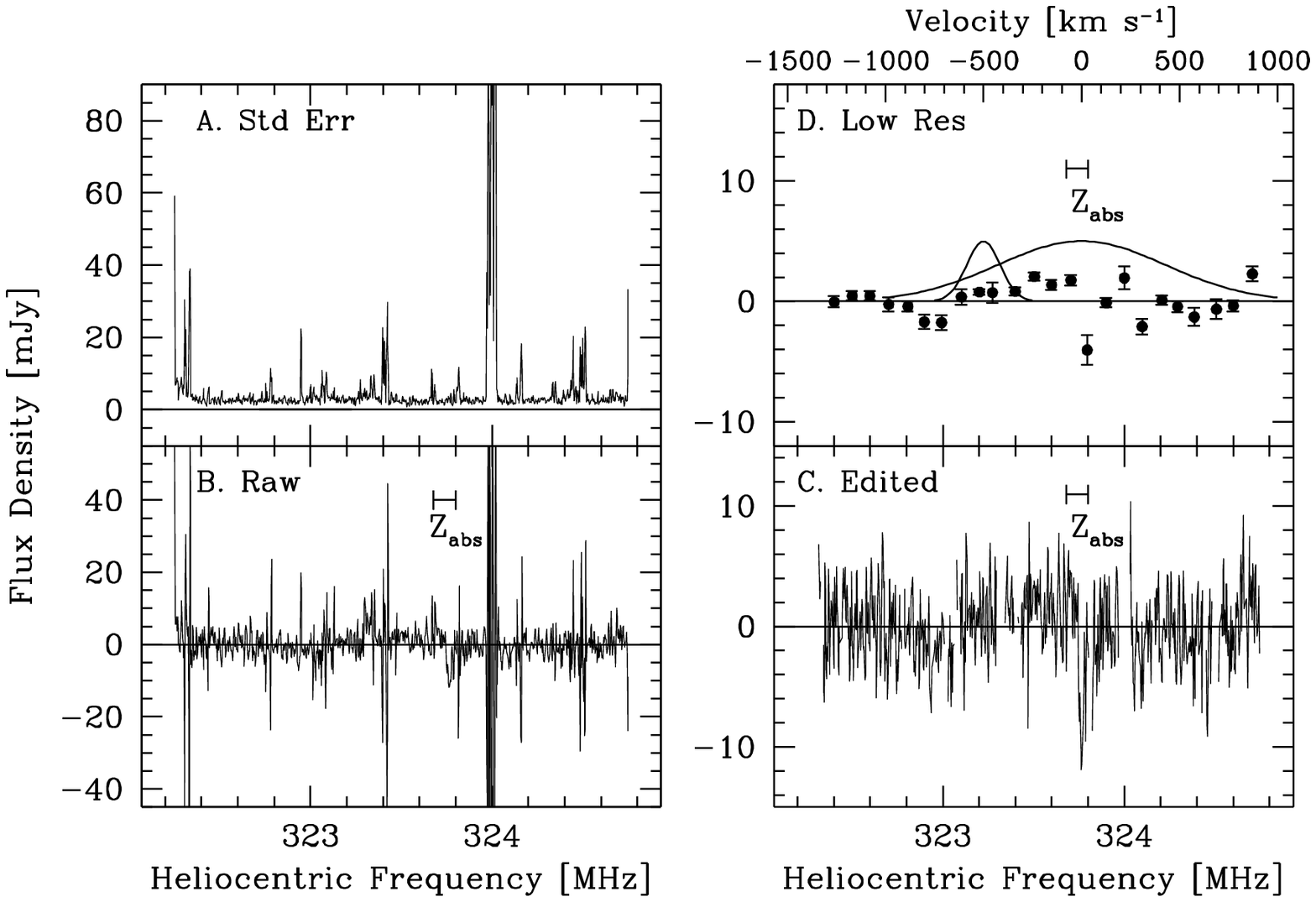}
\figcaption[Briggs.fig1.ps]{Spectrum of MG0201+113. 
{\it A. Top left panel.} Standard error in the
raw spectrum as a function of frequency. {\it B. Lower left panel.} The
mean raw spectrum. The redshift for the optical lines
(White, Kinney \& Becker 
1993) is marked by an error bar to indicate
uncertainty in $Z_{abs}$. {\it C. Lower right panel.} Edited spectrum 
after RFI excision. {\it D. Top right panel} Edited spectrum averaged to
100 kHz resolution. The Gaussian profiles represent $10^{13}\msolar$ of HI
with 200 km~s$^{-1}$ FWHM and $5{\times}10^{13}\msolar$ of HI
with 1000 km~s$^{-1}$ FWHM.
}
}
The ``uncorrupted,'' gain-corrected, 
difference spectra for MG0201+113
were averaged to form one spectrum for each
of the 9 days on which observations were made. 
The weighted average of the 9 nine days is shown in  Figure 1b
after continuum subtraction with a linear baseline.
The standard error of the mean for each spectral point
is shown in  Figure 1a; it was computed from the scatter of the nine
days data about the mean.  The standard error in the final spectrum
was expected to be $\sim$2.8 mJy based on the sensitivity of the system and
the net integration time. 
The variable nature of RFI signals causes afflicted channels
to stand out in the plot of standard error, and this provides a basis for
the automatic excision of corrupted channels from the plot in 
Figure 1c whenever the standard error exceeds a threshold 
of 4.2 mJy, a value which
was chosen by inspection to cut off the non-Gaussian tail 
of the distribution of standard errors. This removes the worst of the RFI
but does not guarantee that all of it has been recognized. Regions
have also been omitted at 323.01-323.03 and 323.30-323.33 MHz, where there is
clear evidence for contamination in Figure 1a and 1b, even though
they failed to be rejected automatically.

Both Figure 1b and 1c show an absorption feature
at a frequency consistent with the redshift determined from optical
spectroscopy of the narrow metal absorption lines associated with the damped
Lyman-$\alpha$ line.
Figure 2 displays 
a 0.5 MHz spectral band encompassing the absorption-line frequency
along with Gaussian profiles that have been fitted 
to the feature. One of the fits relies on the
linear baseline fitted to the entire edited spectrum and produces a line
depth of 11.1$\pm$1.3~mJy. A linear baseline fit to the restricted
spectral region
shown in Figure 2, with the exclusion of the absorption line itself,
has a weak slope across the band and produces a slightly deeper line of
11.6$\pm$2.0~mJy. Both fits indicate a frequency centroid of 
323.764$\pm$0.005 MHz and a width of 25$\pm$5~kHz (FWHM). 
These parameters correspond to a heliocentric
redshift of 3.38716$\pm$0.00007 and a 
velocity width of $\Delta V = 23\pm$5~km~s$^{-1}$. Considerable non-Gaussian
noise remains in the spectra of Figures 1 and 2, due to the 
low level RFI that is scattered throughout 
the band.  Thus, systematic errors
may add to the statistical uncertainties that are determined from the non-linear
least squares fits to the absorption feature.  The integrated on-source and
off-source passband spectra were checked to be sure that the absorption
feature did not arise from a slight excess of weak RFI in the off-source
spectra. The results from the 9 days were also compared to ensure that
the strength and Doppler corrected frequency of the feature is consistent
with a celestial origin.

{
\plotone{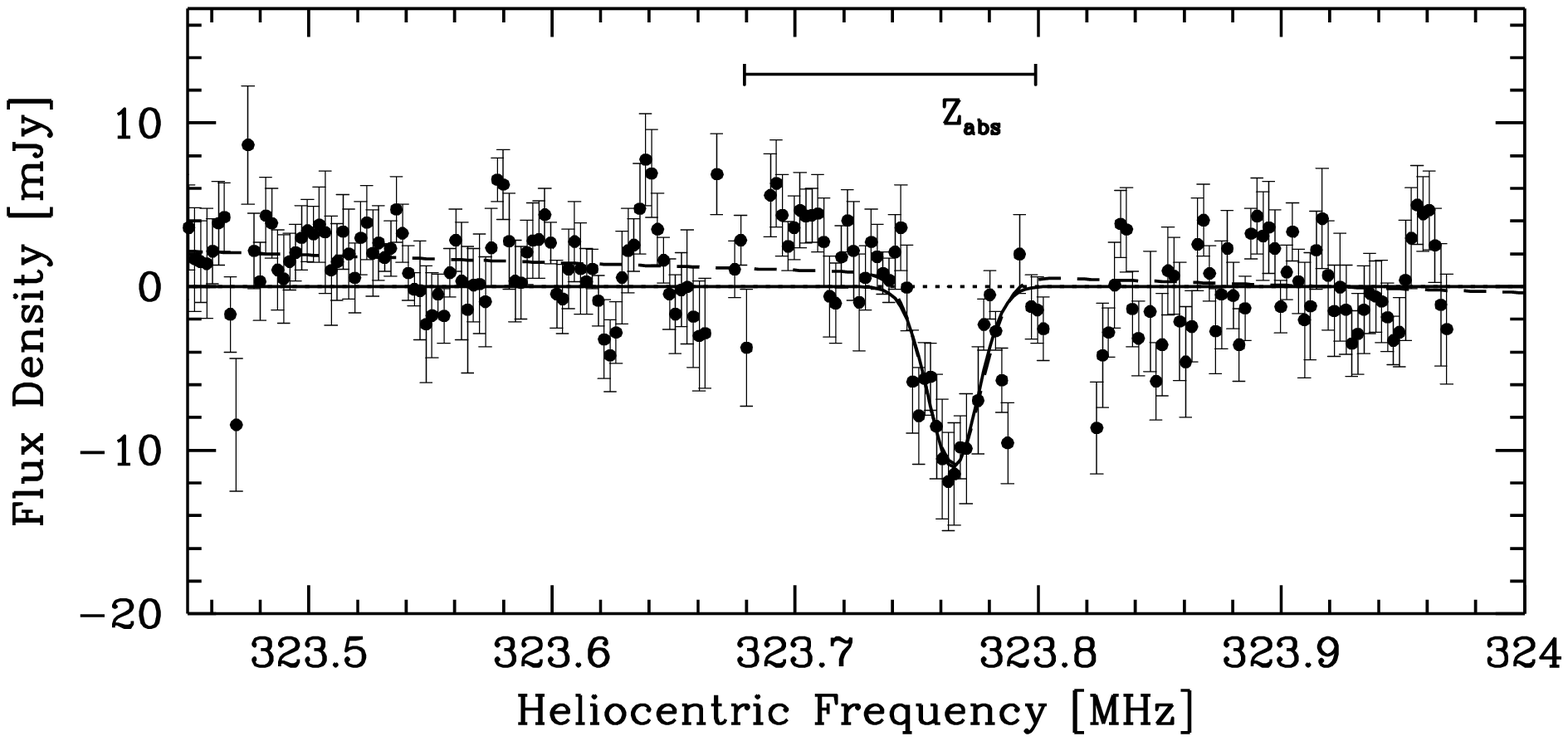}
\figcaption[Briggs.fig2.ps]{Spectrum of MG0201+113 showing the
region of the 21--cm absorption line
at 323.764 MHz. Error bars represent the standard error for each
spectral point.  The solid line shows the fit of a Gaussian line shape
to the absorption line using the linear baseline fitted to the entire
2.5 MHz spectrum after RFI editing. A dashed line, rising to about
$+$2~mJy at the left side of the plot a the baseline
fitted to the limited portion of the spectrum shown in this plot excluding
the region around the absorption line. 
}
}

\section{The VLA Observations}

The source MG0201+113 was observed in D-array on 10 November 
and 7 December 1989
for 1 hour and 1.5 hour respectively. The rms noise reached was only
of order 20 mJy  and was not expected to be sufficient to detect any
absorption. 
The purpose of these observations was to test which observational set-up
was best suited for the project and to obtain a wide--field map over
about 15 degrees. Note that the VLA, because of its antenna lay--out,
collects visibilities in a 3--dimensional volume rather than the usual
2--dimensional {\it uv--}plane (see Perley \markcite{P1} 1988). 
This causes sources away from the
field center to become increasingly distorted. This distortion is
baseline dependent. Therefore, in order to be able to remove strong
sources within the field of view from our high resolution
observations, a low resolution map was essential to locate those
sources.

The high resolution observations were made with the VLA in its
B--configuration. We had six observing sessions, lasting 8 hours each,
of which 6 hours were spent on target, on 23 September and 24--28
October 1991. We used 3C48 as absolute flux calibrator, phase-- and
bandpass calibrator. The observations took place during
reconfiguration from BnA to B-array, except for the 23rd September
observations which were done while the array was mostly still in
A--array. One of the antennae was located on one of the far stations
on the eastern arm. Although it joined in the observations, it's
disproportionally long baseline made it difficult to calibrate and we
decided to remove this antenna from the data. Also for the 23rd
September observations we had difficulties reaching the same level of
accuracy as in the other five runs. In the end we decided not to use
these data, resulting in a total of 30 useful hours on source.

The observations were made at a fixed frequency. Known VLA
interference spikes were as far as possible centered in the middle of
a channel to minimize ``ringing''. The October data were obtained
within one week at a time when MG0201+113 transited at midnight
and no correction for a change in Doppler shift due to
the motion of the Earth around the Sun was necessary since the
frequency  offset was less than 2 kHz. In order to
correct the September data to Heliocentric velocities, the data had to be
shifted by 16.5 kHz.

The observations were done using a spectral line correlator mode in
which the pre--amplified signal is sent through two different
back--end filters. In this way we obtained two sets of data; we
covered 0.781 MHz at a resolution after Hanning smoothing of 12 kHz using
64 channels. Simultaneously, we observed a 0.195 MHz range at 3 kHz
resolution, again using 64 channels. We will concentrate in what
follows on the lower resolution data.

Because of locally generated interference, which is most prominent at
multiples of 12.5 MHz, 5 MHz and 1 MHz, with lesser spikes at 100 kHz
intervals, we had to discard the standard continuum channel, which is
the average over the inner 75\% of the band, and had to create a new
continuum channel based on those channels which were free from
interference and outside of the range where absorption was
expected. From the viewpoint of an interferometer, interference
appears to come from the direction of the North Pole. RFI affects
synthesis maps with characteristic patterns corresponding to 
regions of the  {\it uv--}plane  where phase winding is
low, i.e. along the {\it v--}axis. Therefore, a strip which was
.05 k$\lambda$ wide was discarded from the data.

After these steps, the data were edited and calibrated as usual within
AIPS. Because the observations were done during reconfiguration, some
antennas were either missing from the array, or had such uncertain
baseline positions that we decided to delete them from the
observations. On average, 22 antennae were present during each run. At
327 MHz the ionosphere introduces rapid phase variations. We used
self--calibration (one phase self--calibration cycle, followed by one
cycle correcting both phases and amplitudes) to improve upon the standard
calibration. We used one main field and sixteen subfields within the
task MX to incorporate as many of the brightest sources in the field
as possible for the self--cal.

A critical step in the calibration is the bandpass correction. At
327 MHz, our target field contains some 15 Jansky of flux density. In
order for the final accuracy in the spectra to be determined by the
thermal noise (or confusion limit) rather than by the rms in the
determination of the bandpass correction, the bandpass calibrator
needs to be exceedingly bright. The source 3C48 is about 45 Jy, or $3
\times$ the integrated flux in the field. By observing it for about
one third of the time as the target field, we ensured that we were not
limited by the spectral line dynamic range. Regarding the bandpass, we
know that due to slight temperature variations within part of the
waveguide system, the VLA bandpasses are time variable. A standing
wave pattern with a roughly 3 MHz period moves as a function of time
across the band (both in phase and amplitude). 
To
minimize its impact on our observations, we observed 3C48 every half
an hour between observations on our target and applied the baseline
solution which was closest in time to each target observation.

As was mentioned above, the D--array data were used to map a large, 15
degree, field around MG0201+113. This map showed little distortion and
some 40 sources with a total flux density (not corrected for beam
attenuation) of about 10 Jy were located. Using the AIPS program MX we
mapped small fields, accurately centered on each of these sources to
obtain distortion free maps, and clean components, for all of
them. Because of a limit to the number of fields MX can image at a
time, half the sources, the strongest ones, were located first and
subtracted from the {\it uv--}data using UVSUB. After this the second
set of sources was dealt with.

The VLA observations measure the continuum flux density of
MG0201+113 to be 290$\pm$5 mJy at 330 MHz.

Note that it isn't possible to make a spectrum at the location of
MG0201+113 before any of the sources has been subtracted. The noise in
the maps is dominated by sidelobes of the many sources in the field
rather than thermal noise. At the time the data reduction was
performed, there was no task which could perform distortion-free
imaging of VLA B--array data and the number of fields which could be
cleaned simultaneously was limited as well. After subtraction of the
sources in the {\it uv--}plane as outlined above, we reached the
expected thermal noise in an individual channel map of 6--7 mJy
beam$^{-1}$. It should be noted, though, that still some non--Gaussian
component to the noise is present. When averaging about 48 channels,
the noise doesn't decrease below 3  mJy beam$^{-1}$. Luckily, the
noise characteristics between individual days seem to be uncorrelated
and averaging maps from the October run resulted in spectra with an
average rms noise per channel of 2.8 mJy beam$^{-1}$.

Before creating these final spectra we clipped the visibility data at
6 sigma per individual baseline in order to 
remove random RFI spikes.
Also, we removed the residual continuum
emission 
from the final image cubes with the program IMLIN which
fits a first order baseline (offset and slope) to the data.

These observations were performed in the same period as those by Uson
et al. \markcite{U1} \markcite{U2}
(1991a,b) who 
identified a problem with the setting of
the instrumental delays 
at 327 MHz, causing a subtle instrumental effect to contaminate their
result (Uson, priv. comm.). We checked our observations for a similar
effect. However, our bandwidth  was much narrower than theirs and
any delay errors, visible as a slope in the phases of the bandpass
calibrator, were small and were corrected to first order in the
calibration process.

J. Uson, independently and using the program UVLIN to subtract
continuum sources from the data cubes and remove interference spikes,
re--reduced a few of our observations. He reached broadly similar
noise levels, and the final spectra and maps were virtually identical.

{
\plotone{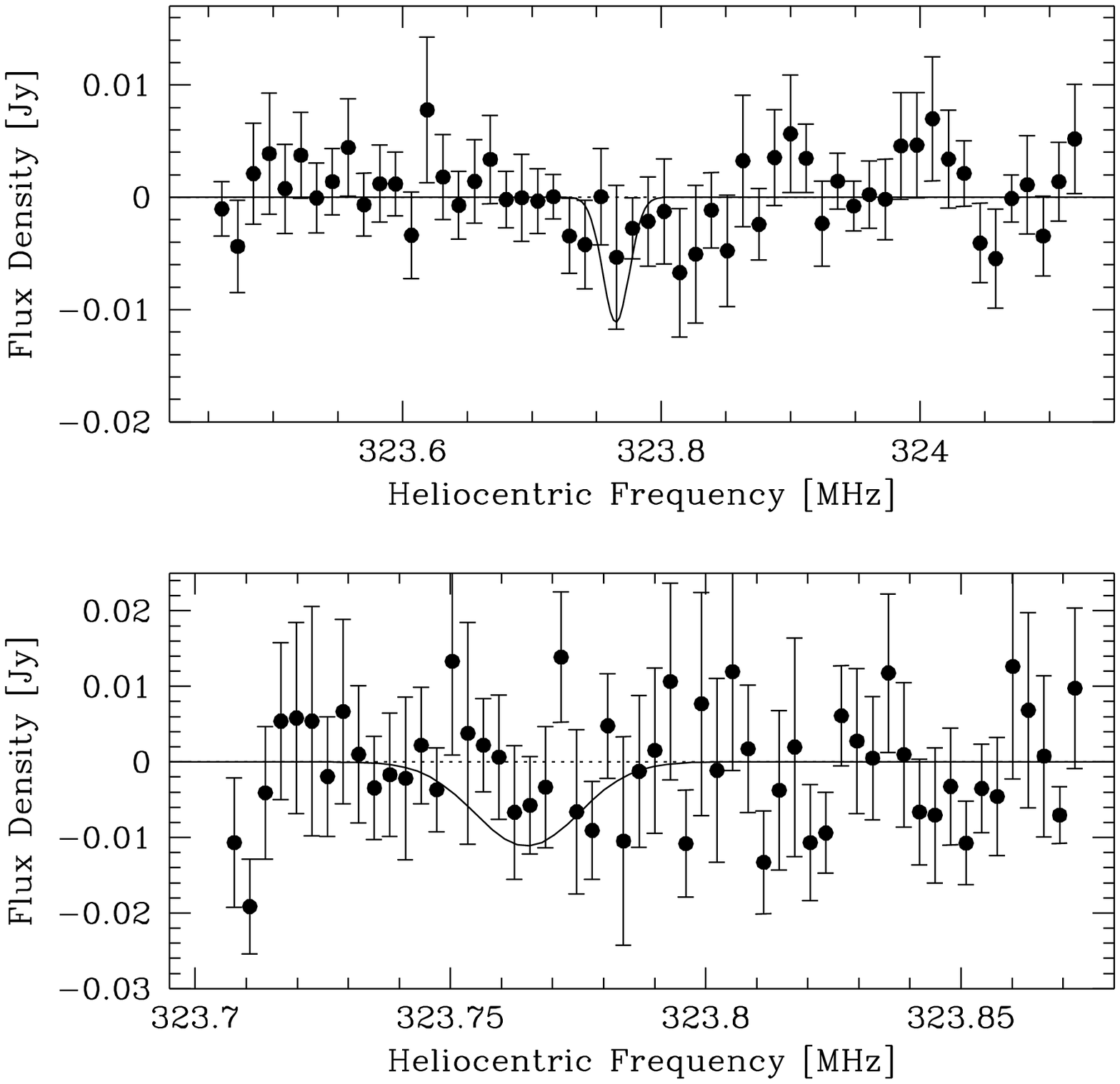}
\figcaption[Briggs.fig3.ps]{VLA Spectra. The upper 
panel shows the low spectral resolution 12.2 kHz
data. The lower panel shows the 3.05 kHz resolution spectrum.
The gaussian fit to the absorption-line measured in the Arecibo spectrum 
is drawn as a solid curve.  Error bars were computed from the internal
consistency of the spectra obtained from the 5 days of observation averaged
to obtain the final spectra.
}
}

The VLA spectrum is shown in Figure 3a. The error bars in this plot
have been determined from the scatter of the measurements for the five
days. The uncertainties computed in this way are expected to have a
spread with standard deviation of ${\sim}6.7/\sqrt 5 = $3.0 mJy,
although there are local variations. No significant absorption is seen
in the VLA spectrum toward MG0201+113 over the optical redshift range
measured at a (local) $3\sigma$ level of 11 mJy beam$^{-1}$, although the
upper panel of Figure 3 shows there is a tendency for the broad-band
data points in the vicinity of the Arecibo line to fall below the zero
line and for points away from the center of the spectrum to lie above.
The noise in the VLA spectra is higher than that obtained at Arecibo,
and the two results are therefore in marginal agreement. However,
neither the VLA nor Arecibo spectra are consistent with the Westerbork
measurement of a line depth of 30 mJy, a width of 15 kHz, and a
heliocentric frequency of $323.7766 \pm 0.002$ MHz. The VLA
measurement at this point is 6.1$\pm$6.2 mJy. Depending on which rms
noise figure one prefers, the line depth measured at Westerbork
would have corresponded at the VLA to a  four to seven $\sigma$ result.

We also extracted spectra towards some of the other strong sources in the
field, since the Arecibo beam has a 16$'$ beam that would include the
spectra of a number of sources at the center of the VLA fields.
No absorption was seen in any of the spectra over the frequency
range covered. The channel maps for the five days were co--added and
were inspected for low--level extended emission. Again, no emission at
a level of 20 mJy over areas of typically 6 arcmin was seen.

Because the WSRT reported (de Bruyn et al. \markcite{B2} 1996) a line
which was not resolved at 9 kHz resolution, we reduced the high
frequency resolution data as well, using the same procedure as
described above. As expected the high resolution spectra have a noise
which is twice that of the low resolution data. The final spectrum is
shown in Figure 3b. No feature was seen at the level reported by the
WSRT, either in the high resolution spectra or in spectra smoothed to
an intermediate resolution. When smoothing even further, to the level
of the low resolution data, basically the same data, but processed
through a different IF system, are recovered.

A comparison of the absorption line shapes measured by the three 
different telescopes is made in Table 1.  The Arecibo and VLA
measurements are consistent with each other but are inconsistent with
the WSRT observation.  An ``integral line strength'' of the sort computed
in the following section for the purpose of estimating the gas
spin temperature is also listed.  These values are similar, since the
WSRT line is both deeper and narrower by factors of approximately three
than the Arecibo measurement; this explains why similar constraints are
placed on spin temperature by de Bruyn et al (1996) and us.
 
%
%

\begin{deluxetable}{lrrr}
\tablewidth{440.0pt}
\tablenum{1}
\tablecaption{Comparison of  Line Parameters at Three Telescopes}
\tablehead{
\colhead{Parameter} & \colhead{WSRT$^a$} &
\colhead{Arecibo$^b$} & \colhead{VLA$^c$}     
}

\startdata
Optical Depth $\tau_{21}$    & 0.085 & 0.037  &  $<.05(23/{\Delta}V)^{1/2}$   \nl
~ ~ at line center & ${\pm}$0.02 & ${\pm}$0.008 & \nl
           &         &      &                 \nl
Line Width (FWHM) ${\Delta}V$& 9  & 23  & --- ~~~~~~~~~~   \nl
~ ~ km s$^{-1}$   & ${\pm}$2 &  ${\pm}$5  &   \nl
                  &         &      &         \nl
Redshift, $z_{abs}$  &  3.38699  & 3.38716  & --- ~~~~~~~~~~  \nl
~ ~ Heliocentric      & ${\pm}$0.00003  & ${\pm}$0.00007  & \nl 
                  &         &      &         \nl
$\int{\tau_{21}dv}$  &  0.77  & 0.85  & $<1.2({\Delta}V/23)^{1/2}$ \nl
~ ~ km s$^{-1}$   & ${\pm}$0.2  & ${\pm}$0.2  & \nl 
\enddata
\tablenotetext{a}{Westerbork data from de Bruyn et al (1996).}
\tablenotetext{b}{Arecibo observations from this paper.}
\tablenotetext{c}{The VLA observation reported in this paper
places upper limits on the optical depth and integral line strength.}
\end{deluxetable}

\section{Discussion}

In principle, a comparison of the column density obtained from
measurement of the Lyman~$\alpha$ absorption-line with the optical
depth measured in the radio can be used to specify the spin temperature of the
absorbing gas, thus indicating the excitation level in this high redshift
absorber. This line of reasoning has been pursued in a number of cases
(Wolfe \& Davis 1979, Wolfe et al 1981, Wolfe et al 1985).
Here, we pursue this argument, pointing out the large
uncertainties  and leading to the conclusion that the result would
be considerable strengthened by further radio studies to define the 
background quasar's radio continuum structure and new high-resolution optical
spectroscopy to better define the complete kinematical model for this absorption
system. Similar arguments have been made by de Bruyn et al (1996) for
this system.

Combination of the observed 21--cm line depth with the continuum strength 
produces
a measure of optical depth at line center of $\tau_{21} = 0.037\pm 0.008$.
Adopting the column density determined from the strength of the
damped Lyman~$\alpha$ absorption line 
(White, Kinney \& Becker \markcite{W2} 1993) 
$N_{HI} = 10^{21.4}$cm$^{-1}$ and 
using the expression for the spin temperature, 
$<T_s> = N_{HI}/(1.823{\times}10^{18}\int \tau_{21}(v) dv)$ 
(cf. Dickey \& Brinks \markcite{D1} 1988), yields an estimate
for $<T_s> \approx 1380$~K.  This temperature is the 
column-density-weighted, harmonic mean
of the spin temperature along the sight line
(Kulkarni \& Heiles \markcite{K1} 1987, Dickey \& Lockman  \markcite{D2} 1990).
$<T_s> \approx 1380$~K is substantially higher than is observed along
lines of sight through the Milky Way Galaxy or M31 (Dickey \& Brinks 
\markcite{D1} 1988),
where $<T_s> \sim$ 200 to 300~K are typical for column
densities $N_{HI}\approx 10^{21}$cm$^{-2}$. 
The estimate for $T_s$ in the $z=3.4$ absorber
does have  uncertainty in addition 
to those attached to $\tau_{21}$ and $\Delta V$;  we estimate the uncertainty
in $N_{HI}$ to be ${\sim}$30\%, which is not enough to alter the
conclusion that $<T_s>$ computed in this way is high.

We have assumed that all the neutral gas that
contributes to the damped Lyman~$\alpha$ profile is confined  to the 
Gaussian feature that we observed in the 21--cm line.  A sharply
contrasting possibility is that the high redshift
system has substantial 
quantities of warm, low optical depth gas that is sufficiently spread
in velocity to escape inclusion in the 21--cm profile. Then, the inferred
$T_s$ for the 21--cm absorber becomes ${\simeq}T_sf_{21}$, where $f_{21}$ is
the fraction of $N_{HI}$ that is contained in the 23 km~s$^{-1}$ 
profile (velocity dispersion $\sigma \approx 10$ km~s$^{-1}$).  Thus,
in addition to the narrow 21--cm absorption component, a complete
description of the kinematics of the system must include
the broader, low column density component whose 
cloudlets are optically thick in the resonance lines of common metals
but optically thin in the 21--cm line. In other systems at lower $z$,
this component has been tentatively
identified with halo gas that encases the dynamically cold, 21--cm absorbing
disk (Wolfe \& Wills \markcite{Wo2} 1977, Briggs \& Wolfe \markcite{Br1}
1983, Briggs et al \markcite{Br2} 1985, 
Lanzetta \& Bowen \markcite{L1}
1992). For these cases, high-resolution optical 
spectroscopy has led to 
the general conclusion that most
of the neutral column density in the damped Lyman~$\alpha$ systems is 
concentrated in the cold, low-dispersion
component (c.f. Wolfe et al \markcite{Wo5} 
1994), and only a tiny fraction of the
neutral gas  is contained in the broad, turbulent (or ``halo'')
component, which dominates in providing the equivalent widths measured for
the metal lines. Thus, $f_{21}$ is likely to be
close to unity, and $T_s \approx 1380$~K
as indicated above.

For this system in absorption against MG0201+113,
the effective velocity spread of the broad component can be estimated
from the equivalent widths of the metal lines observed in the low
resolution (10\AA) spectra of White, Kinney and Becker \markcite{W2}
(1993). We 
performed a curve of growth analysis using the line depths and limits to
depth for several transitions of SiII ($\lambda$1260, $\lambda$1526, 
$\lambda$1808), FeII ($\lambda$1144,
$\lambda$1608), CII ($\lambda$1334), and AlII ($\lambda$1670).  
Although there is considerable 
uncertainty in this analysis, it serves to illustrate the constraints on
the parameters that need to be specified
to better determine the spin temperature. 
A range of two component models is illustrated in Figure 4: the cold
cloud has $\sigma_{21} = 10$ km~s$^{-1}$ and $N_{21}=f_{21}N_{HI}$, and
the ``halo'' component is modeled as a single broad dispersion cloud
with $\sigma_{ha}$ and neutral hydrogen column density 
$N_{ha}=f_{ha}N_{HI}=(1-f_{21})N_{HI}$. Due to the scarcity of information,
a single measure for the metal abundances relative to solar 
abundances ``$X$'' is assumed for all
ions in both components; for example, the total column density of CII 
in the system is 
specified by $N_{CII}= 10^XN_{HI}{\times}[n_C/n_H]_{\odot}$, where
$[n_C/n_H]_{\odot}$ is the solar abundance of carbon relative to hydrogen
as tabulated by Spitzer \markcite{S2} (1978).
This approach also assumes that the carbon associated with the $N_{HI}$
in this layer is predominantly singly-ionized, in keeping with neutral
clouds along lines of sight through the Milky Way Galaxy. The
upper panels of Figure 4 illustrate models in which the 21--cm absorber
dominates the neutral column density; the lack of detection of the 
SiII$\lambda$1808 line enforces a low metal-abundance estimate, and the 
stronger detections
of other metal lines indicates that there is additional, turbulent gas.
The data permit the velocity spread of the turbulent component to
be very broad, although the CII line would be resolved by the instrumental
resolution if the effective $\sigma$ were much over 200~km~s$^{-1}$.
The low resolution spectroscopy is compatible with a wide range of models.
For example,
the metal line strengths are also compatible with much larger fractions of
gas lying in the broad component, as is shown in the lower panels of
Figure 4; this range of models must have much lower metal abundances.

{
\plotone{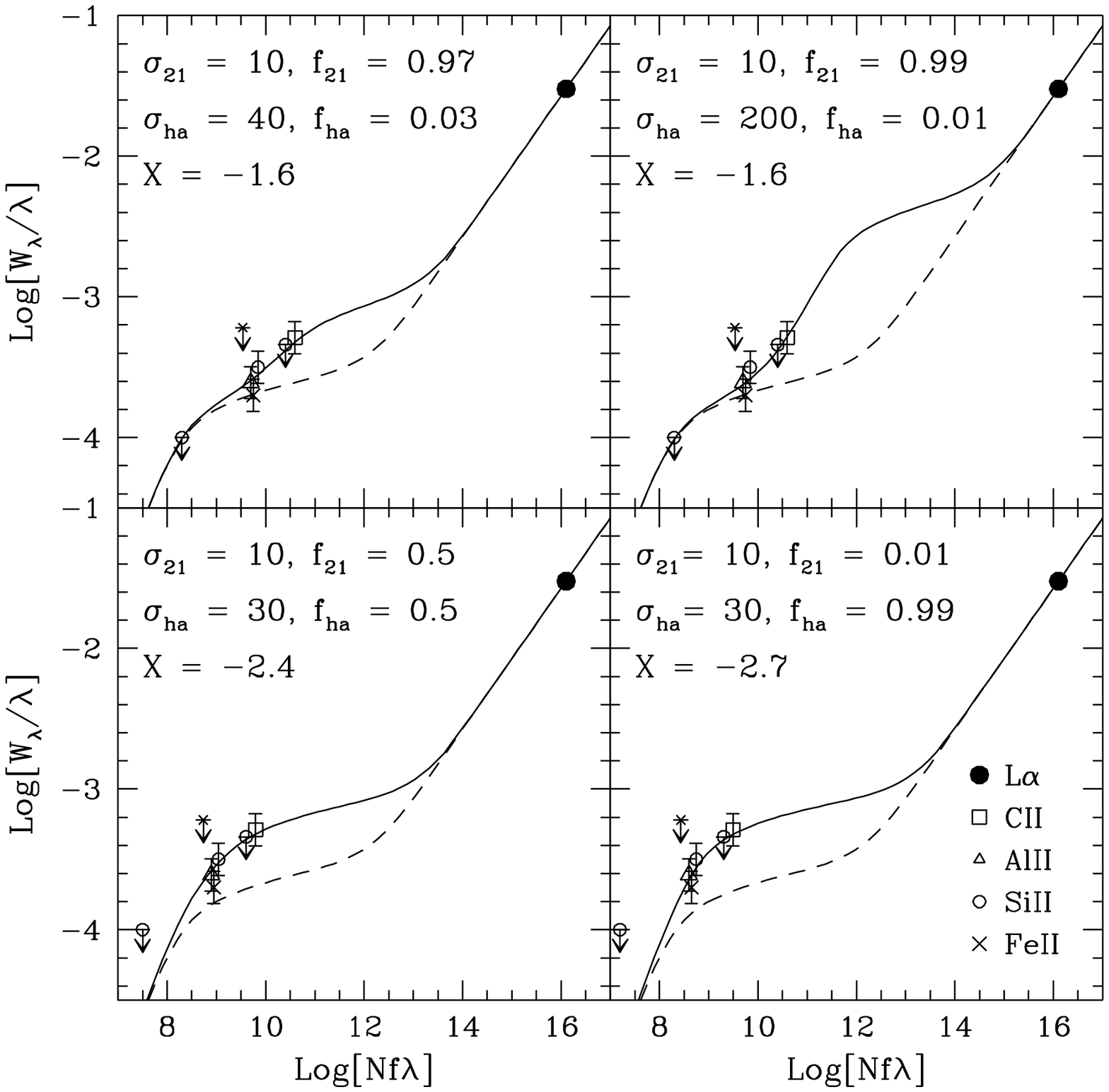}
\figcaption[Briggs.fig4.ps]{Curve of Growth (COG) 
display of the range of two-component
kinematical models that
simultaneously fit both the 21--cm line profile and the ultraviolet lines.
The {\it top panels} illustrate the range of models with the highest 
metal abundances (relative to solar abundances) and lowest fraction of
neutral gas in the broad velocity dispersion, ``halo'' component. The
{\it lower panels} explore the range of models with the large fractions
of gas in the ``halo'' component, yielding the lowest values of metal
abundances.  The dashed COG represents the $\sigma_{21} = 10$
km~s$^{-1}$ 21--cm absorption component. The solid line is the two component
model. The damping portion of the COG is drawn for Lyman~$\alpha$.
The abscissa has units of cm$^{-1}$. 
}
}

The constraints on the relative column densities in the turbulent and
cold components
can be translated into measurements of the spin temperatures for both.
These spin temperatures and the other model parameters from the COG 
analysis are summarized in Figure 5.
The metal abundances from the models range from $10^{-1.6}$ to $10^{-2.7}$
below solar.  The lower limit to the
spin temperature of the turbulent halo component results from inspecting
the spectra in Figures 2 and 3 and noting that an absorption-line three times
broader than the detected line would probably escape detection if its depth
were less than $\sim$5~mJy. Although $f_{21}$ is expected to be close to 
unity, the certainty of the interpretation would benefit from new,
high-resolution optical spectroscopy that can demonstrate that the bulk
of the gas is confined to the 21--cm line velocity width.

{
\plotone{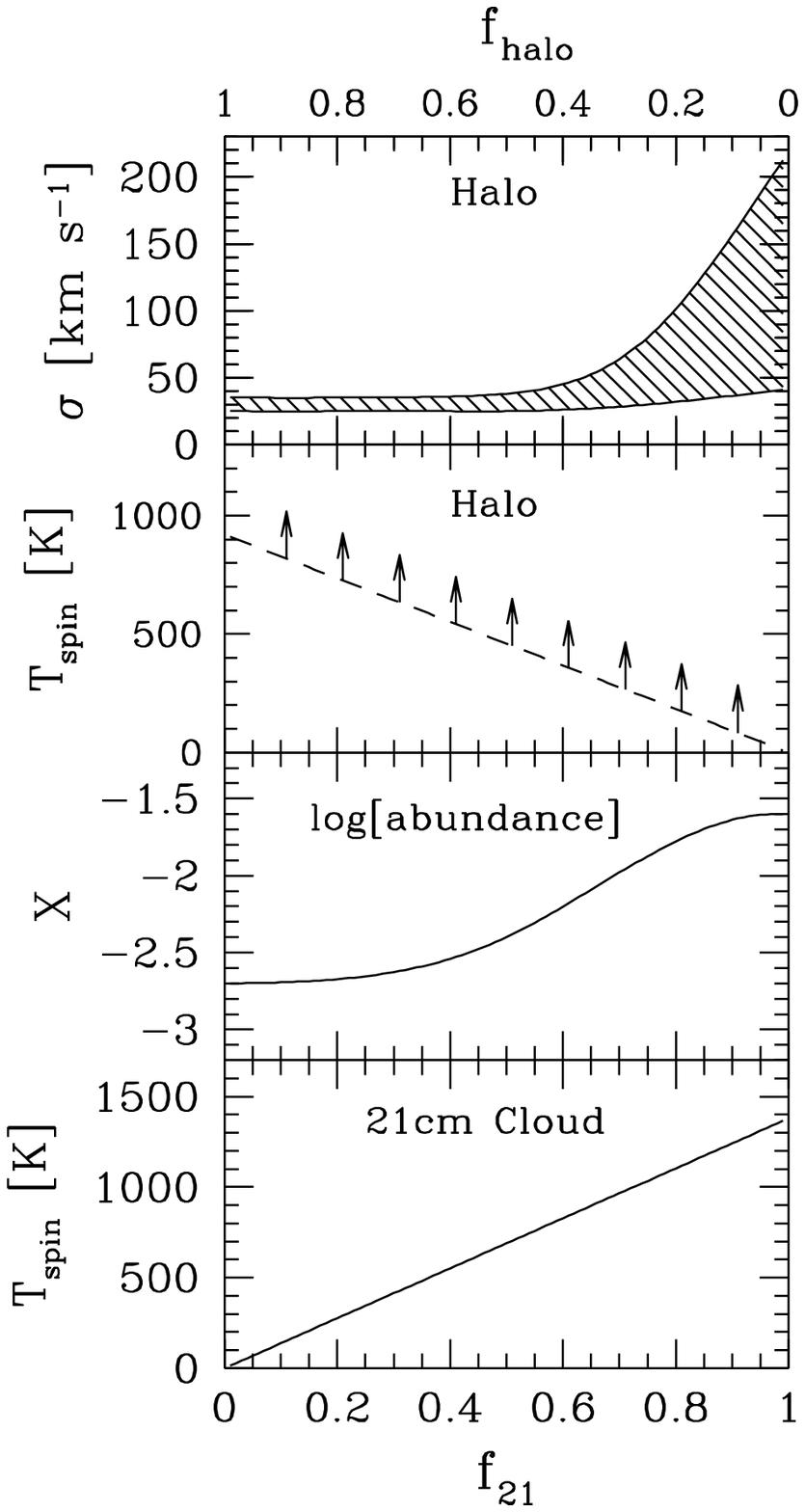}
\figcaption[Briggs.fig5.ps]{Constraints on spin 
temperature as a function of the fraction of
the neutral gas contained in the 21--cm component, $f_{21}$.  The fraction of
gas contained in the broad velocity dispersion ``halo'' component, 
$f_{halo} = 1 - f_{21}$, is plotted across the top horizontal axis.
{\it Upper panels} indicate the range of effective velocity dispersions, 
$\sigma$,
and lower limit to spin temperature, $T_{spin}$, for the ``halo'' component,
determined
from the COG analysis of Figure 4 and the detection limits in the 21--cm line.
The {\it lower-central panel} plots the logarithm of the metal abundance
relative to solar, assuming that it is the same in both components.
The {\it lower panel} plots the spin temperature of the 21--cm line component
as a function of $f_{21}$.
}
}

Further uncertainty in determination of the spin temperature
results from possible differences in the spatial extent of the absorber and
the background radio source,
which may be substantially larger than
the source of optical continuum causing a significant portion of the radio
flux to leak around the thick absorbing cloud 
that creates the damped Lyman~$\alpha$ line. The continuum map at 18cm
wavelength (Stanghellini et al \markcite{S3}
1990) hints that this might be the case
by showing evidence for a weak component
extending to ${\sim}4''$ from the nucleus, although the extended emission
accounts for only a tiny fraction of the total radio flux. 
While MG0201+113 falls in the GHz--Peaked-Spectrum class of radio source,
the integral flux density spectrum for the source shows a flattening
at meter wavelengths (O'Dea et al 1990), indicating 
that at the frequency of the redshifted
21--cm absorption line, the source is dominated by a more extended 
component than the compact optically thick core component that gives
rise to the GHz peak in the continuum spectrum (de Bruyn et al 1996).
High resolution
90cm VLBI observations are capable of defining the continuum source structure, which ultimately
will clarify the interpretation.  If the resulting $<T_s>$ remains high, then
the neutral medium in this absorber is
substantially hotter than the neutral ISM of the Milky Way or M31. This may
indicate that a larger fraction of the total HI 
is in the Warm Neutral phase with
$T \approx 8000$~K (Kulkarni \& Heiles \markcite{K1} 1987), or
alternatively,  the gas in the
Cold phase that we detect in the 21--cm line
may actually be hotter due to the inability of 
the gas to cool itself when the metal abundances are low.

The new Arecibo HI spectroscopy also places 
limits on the emission from large, neutral gas masses that might be associated
with the damped Lyman~$\alpha$ site. The sensitivity is a strong
function of the velocity width of the signal
(cf. Wieringa et al. \markcite{W1} 1992), which would be expected to
range from a few hundred km~s$^{-1}$ for  pancakes (Zel'dovich 1970) 
to more than 1000 km~s$^{-1}$ for analogs to present day galaxy clusters.
Two examples of Gaussian profiles for $\Delta V = 200$ km~s$^{-1}$,
$M_{HI} = 10^{13}\msolar$ and $\Delta V = 1000$ km~s$^{-1}$,
$M_{HI} = 5{\times}10^{13}\msolar$ ($H_o$ = 100 km~s$^{-1}$~Mpc$^{-1}$
and $\Omega_o = 1$) are plotted in Figure 1d for comparison with
the data of Figure 1c, after averaging to 100 kHz wide spectral bins.
Only a single linear baseline has been subtracted from the entire 2.5 MHz
band. 
The choice of the narrow 2.5 MHz band for this experiment limits
would prevent the detection of 
$\Delta V$ 
$\lower .7ex\hbox{$>$}\atop \raise .2ex\hbox{$\sim$}$ 1200 km~s$^{-1}$.
Clearly, systematic effects, such as a weak, characteristic 
``standing wave'' and low level RFI, dominate in limiting the sensitivity
to broad signals. On the other hand, signals as large as these would 
be noticeable.  The limits are still more than a factor of 10 higher than
Wolfe's \markcite{Wo4}
(1993) prediction for the amount of HI associated with damped
Lyman~$\alpha$ sites.

\acknowledgments

The authors are grateful to members of the
Arecibo Observatory staff, particularly 
E. Castro, B. Genter, A. Noya,  and J. Rosa, for their
improvements to the receiving system as well as the 
Operations Group for their efficient assistance with the
observations.
We are grateful to J. Uson for sharing his expertise with us and for
his valuable advice at various stages of the data reduction. 
The National Astronomy and Ionosphere Center is
operated by Cornell University under contract with the National Science
Foundation. 
F.B. is grateful to the Arecibo Observatory and the Institute for 
Advanced Study for their hospitality during the periods when this
work carried out.
This work has been supported by NSF Grant AST91-19930.

\end{document}